\documentclass[a4paper,11pt]{article}
\usepackage{pos}
\usepackage{subfiles}

\title{The Science Alert Generation system of the Cherenkov Telescope Array Observatory}
 \ShortTitle{The CTA Science Alert Generation system}

\manuallySeparateAuthors
\author*[a]{A. Bulgarelli,}
\author[b]{ S. Caroff,}
\author[a]{ A. Addis,}
\author[b]{ P. Aubert,}
\author[a]{ L. Baroncelli,}
\author[a]{ G. De Cesare,}
\author[a]{ A. Di Piano,}
\author[a]{ V. Fioretti,}
\author[b]{ E. Garcia,}
\author[b]{ G. Maurin,}
\author[a]{ N. Parmiggiani,}
\author[b]{ T. Vuillaume,}
\author[c]{ I. Oya}
\author[d]{ and C. Hoischen}
\author[]{for the CTA Observatory\footnote{https://www.cta-observatory.org/}.}

\affiliation[a]{INAF/OAS Bologna, Via P. Gobetti 93/3, I-40129 Bologna, Italy}

\affiliation[b]{LAPP, Univ. Grenoble Alpes, Univ. Savoie Mont Blanc, CNRS, 9 chemin de bellevue, 74940 Annecy, France}

\affiliation[c]{CTAO gGmbH, Saupfercheckweg 1, 69117 Heidelberg, Germany}

\affiliation[d]{Institut für Physik und Astronomie, Universität Potsdam, Karl-Liebknecht-Straße 24/25, 14476 Potsdam, Germany }

\emailAdd{andrea.bulgarelli@inaf.it}
\emailAdd{caroff@lapp.in2p3.fr}

\newcommand{\abr}{}
\newcommand{\abrb}{}

\abstract{The Cherenkov Telescope Array (CTA) Observatory, with dozens of telescopes located in both the Northern and Southern Hemispheres, will be the largest ground-based gamma-ray observatory and will provide broad energy coverage from 20 GeV to 300 TeV. The large effective area and field-of-view, coupled with the fast slewing capability and unprecedented sensitivity, make CTA a crucial instrument for the future of ground-based gamma-ray astronomy. To maximise the scientific return, the array will send alerts on transients and variable phenomena (e.g. gamma-ray burst, active galactic nuclei, gamma-ray binaries, serendipitous sources). Rapid and effective communication to the community requires a reliable and automated system to detect and issue candidate science alerts. This automation will be accomplished by the Science Alert Generation (SAG) pipeline, a key system of the CTA Observatory.
SAG is part of the Array Control and Data Acquisition (ACADA) working group. The SAG working group develops the pipelines to perform data reconstruction, data quality monitoring, science monitoring and real-time alert issuing during observations to the Transients Handler functionality of ACADA. 
SAG is the system that performs the first real-time scientific analysis after the data acquisition. The system performs analysis on multiple time scales (from seconds to hours). \abrb{SAG must issue candidate science alerts within} 20 seconds from the data taking and with sensitivity at least half of the CTA nominal sensitivity. These challenging requirements must be fulfilled by managing trigger rates of tens of kHz from the arrays. Dedicated and highly optimised software and hardware architecture must thus be designed and tested. In this work, we present the general architecture of the ACADA-SAG system.}

\FullConference{37$^{\rm{th}}$ International Cosmic Ray Conference (ICRC 2021)\\
		July 12th -- 23rd, 2021\\
		Online -- Berlin, Germany}

\begin{document}
\maketitle

\section{Introduction}

A science alert is a communication \abr{within} the astrophysical community to share the information that a transient phenomenon (e.g. AGN's gamma-ray flare, gamma-ray bursts (GRBs), gravitational waves \cite{patricelli}, galactic transients \cite{alicia}) is observed. In the multi-wavelength and multi-messenger context, this sharing through specialised communication networks \abr{enables the coordination of different observatories, allowing to understand the nature of the variable phenomena. To achieve this purpose, the transient phenomena must be followed in real-time at a short timescale. Furthermore, if observatories identify a transient, it is important to act quickly to repoint the telescopes and to analyse the data in an automated and fast way to confirm the transient event and for its follow-up}. In this context, quick reaction time and fast data analysis and the notification of information with science alerts through communication networks allow to maximise the scientific return of an observatory. 

 Current Imaging Air Cherenkov Telescopes (IACTs) (H.E.S.S. \cite{hess}, MAGIC \cite{magic}, and VERITAS \cite{veritas}) are limited in sensitivity for detection on a short timescale (from tens to hundreds of second). Still, the Very High Energy (VHE) domain has a great potential in unveiling both galactic (e.g. binaries) and extra-galactic (e.g. GRBs and active galactic nuclei) variable sources. The Cherenkov Telescope Array \abr{(CTA)} Observatory will fill this gap  \cite{cta}, with tens of telescopes located in both the northern and southern hemispheres. CTA will be the largest ground-based gamma-ray observatory and will provide broad energy coverage from 20 GeV to 300 TeV. The large effective area and field-of-view, coupled with the fast slewing capability and unprecedented sensitivity at short timescales \cite{fioretti}, make CTA a crucial instrument for the future of ground-based gamma-ray astronomy. The final sensitivity will enable the detection and monitoring of gamma-ray transients in the VHE domain \cite{bosnjak}, which is one of the Key Science Projects of the observatory \cite{cta}. CTA can operate in a wide range of configurations thanks to many telescopes, enabling observations with multiple sub-arrays targeting and simultaneous monitoring of different objects or energy ranges. With its large detection area, CTA will resolve flaring and time-variable emission on sub-minute time scales.
 
 The CTA unprecedented scientific performance places the Observatory as a reference for detecting and monitoring gamma-ray transients at the peak of the time-domain astronomy era. Rapid and effective communication to the community requires a reliable and automated system to detect and issue candidate science alerts. This automation will be accomplished by the \textit{Science Alert Generation} (SAG) pipeline, a key system of the CTA Observatory that, employing real-time analysis pipelines running on-site with the telescopes, will be able to generate candidate science alerts within $20 \: s$ from the last acquired event, with a maximum telescope positioning time of 90 s in response to external or internal triggers. Candidate science alerts will be sent to the \textit{Transients Handler}, another key system of CTA, that will evaluate the results of the SAG to generate the final science alert to the community \abrb{within $5 \: s$ of receiving it. If we consider also $5 \: s$ required to acquire data, CTA is capable of issuing science alerts with a total latency of less than $30 \: s$.  }
 
%CTA will be capable of following up Gravitational Waves event candidates over the required large sky area (also 1000 deg2) with sufficient sensitivity to detect short gamma-ray bursts, which are thought to originate from compact binary mergers.

\section{General architecture}
The \textit{Science Alert Generation} (SAG) system is part of the Array Control and Data Acquisition (ACADA) system \cite{Igor} of CTA. SAG is the system that performs the first real-time scientific analysis during the \abr{CTA} data acquisition to perform data reconstruction, data quality monitoring, science monitoring and real-time candidate science alert issuing during observations to the Transients Handler functionality of ACADA. 

\begin{figure*}[!htb]
	\centering
	  \includegraphics[width=\textwidth]{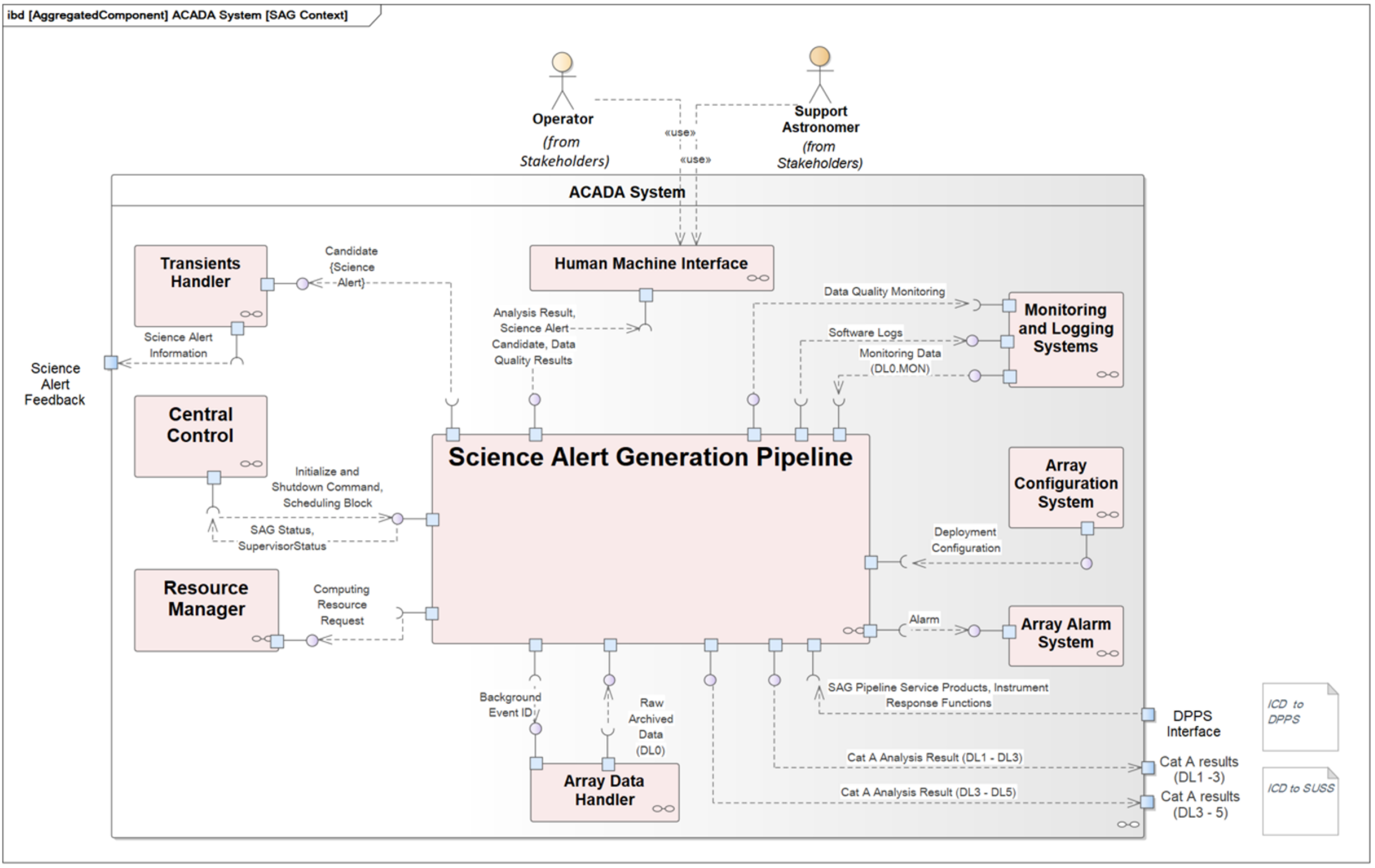}
	\caption{Context diagram of the SAG. The diagram uses the UML component notation, and the basic entities are the software components, depicted as pink boxes. An UML component is an ACADA sub-system. Blue squares are “ports” that identify the interfaces of the systems. Those ports in the boundary of the ACADA identify the external interfaces of the system, and those in the components the internal interfaces. The dashed lines show the flow of data elements and the arrows the direction of the flow.}
	\label{fig:sag}
\end{figure*}

A general view of the SAG context is reported in Fig. \ref{fig:sag}. SAG shares interfaces with many ACADA sub-system, \abr{as} described in \cite{Igor}. In addition to the already cited \textit{Array Data Handler} (to receive the telescope data and stereo event trigger information, \abr{i.e.} DL0) and \textit{Transients Handler}, another important sub-system is the \textit{Central Control} for the execution of the scheduling blocks provided by the \textit{Short-Term Scheduler}, sending corresponding commands to the telescopes and configuring ADH, SAG and \textit{Transients Handler} for data acquisition and analysis for each sub-array. In addition, SAG shall receive environmental information and night sky background level information in real-time. A \textit{Human Machine Interface (HMI)} will provide a comprehensive view of the status of observations to the operators and support astronomer located in the control room of the CTA installation. The \textit{Monitoring and Logging Systems} will collect monitoring data from SAG.

\abr{The Alma Common Software (ACS) framework \cite{chiozzi} is the basic middleware of the ACADA sub-systems. It allows ACADA components to run in the on-site data centre and interact with each other using this framework, providing a common way to define interfaces.}

The main SAG requirements can be summarised as follow:
\begin{enumerate}
 \item 	a high input data rate received from the \textit{Array Data Handler (ADH)} sub-system: SAG must handle 0.3 Gbps assuming $5\%$ of pixels survived zero-suppression, with a maximum data rate of 9 Gbps in case no zero-suppression is applied, with a corresponding event rate of tens of kHz;
 \item \abrb{response time of $20 \: s$ for the generation of candidate science alerts; }
 \item  availability of $95\%$ during the observation time;
 \item  flexibility in terms of scientific targets, science tools and analysis methods required to work with different reconstruction, data quality, and scientific analysis algorithms; 
 \item 	capability to work with different array configurations and with many sub-arrays in parallel;
  \item  prioritisation of urgent scientific targets. 
\end{enumerate}

\begin{figure*}[!htb]
	\centering
	  \includegraphics[width=\textwidth]{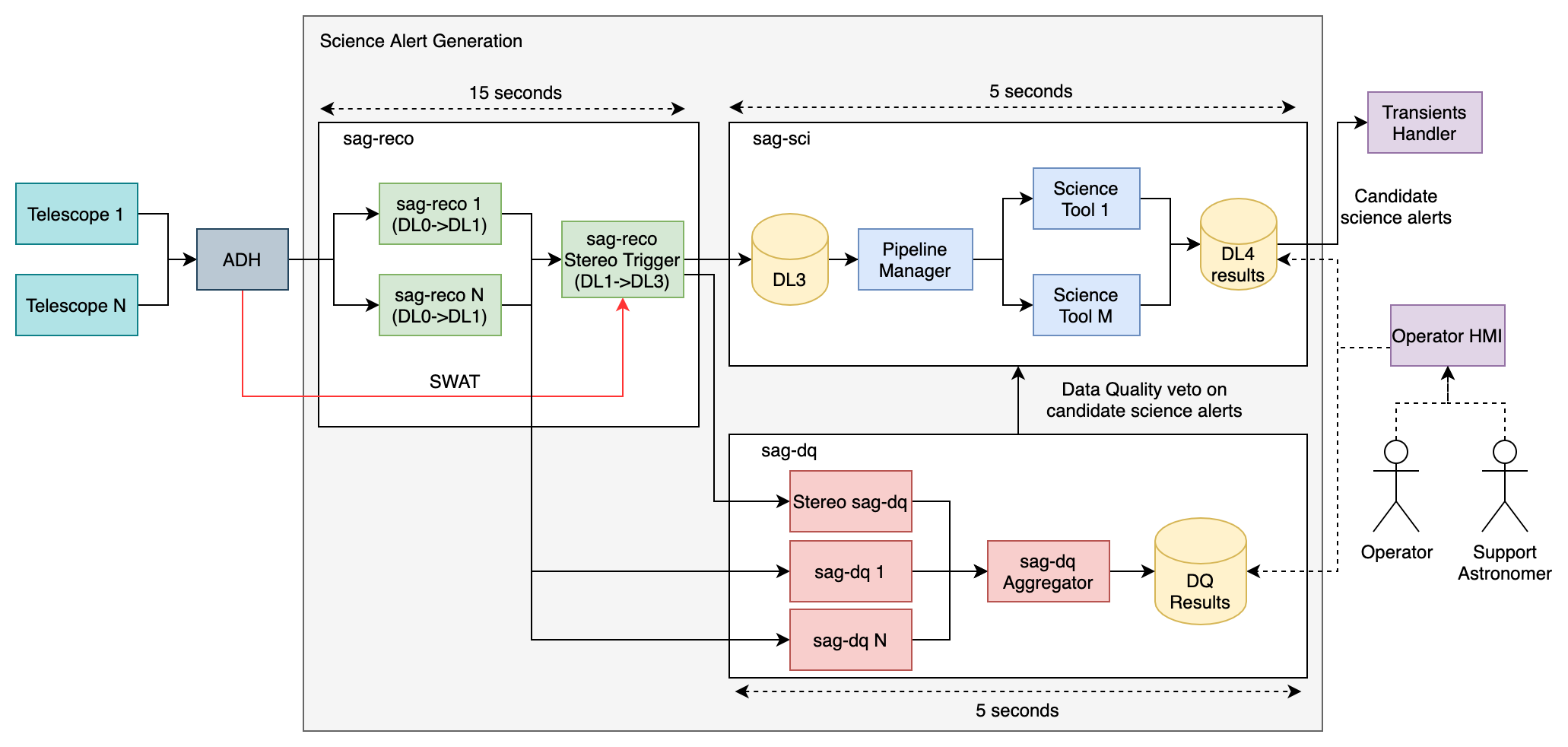}
	\caption{SAG pipelines and dataflow.}
	\label{fig:sag2}
\end{figure*}

\abr{The  SAG pipelines (Fig. \ref{fig:sag2}) are}
\begin{enumerate}
\item 	\textbf{sag-reco}: the Low-Level Reconstruction  pipeline \abr{for fast reconstruction} to process individual DL0 events and produce the corresponding DL3 (the gamma-ray like event list)) data with a latency of less than $15 \: s$. sag-reco  \abr{provides} a library or software component to integrate data-cubes (DL0) into images, then clean images and extract \abr{Hillas} parameters from Cherenkov camera images \abr{ (DL1 data), that are  delivered to other SAG pipelines}. 
\item 	\textbf{sag-dq}: The SAG system also performs an online data quality analysis to assess the instruments’ \abr{status} during the data acquisition \abrb{with a latency of less than $5 \: s$}: this analysis is crucial to confirm good gamma-ray sources and transient detections. 
\item   \textbf{sag-sci}: The High-Level analysis scientific pipelines \abrb{is able to execute scientific analysis and generate candidate science alert  with a latency of less than $5 \: s$}. This is a sub-system  based on a framework designed to simplify the development of real-time scientific analysis pipelines \cite{parmiggiani1}. Using this framework, developers and researchers can focus more on the scientific aspects of the pipelines and integrate existing science tools, providing a common pipeline architecture and automatism. The framework can be easily configured with new or existing science tools \cite{dipiano} implemented in different programming languages. The pipelines are able to execute all the analyses automatically after the initial setup. The scientific analyses are performed in parallel and can be prioritized.
\end{enumerate}

\abrb{The $5 \: s$ latency of \textbf{sag-dq} goes in parallel to the latency of \textbf{sag-sci}. Thus, the total SAG latency is $20 \: s$.}

In addition, a SAG Pipeline sub-array Supervisor is an ACS  component that supervises the operations of the SAG pipelines associated with a sub-array and provides the general interface with the rest of the ACADA system.  When a sub-array enters in observing state, the SAG Supervisor  starts and monitor the related SAG pipelines. When a sub-array stops the observation, the SAG Supervisor  informs SAG pipelines that  shut down data processing in an automatic and controlled way. The SAG Supervisor  sends monitoring information from all running sub-array pipelines to the ACADA \textit{Monitoring and Logging Systems}.

The functional diagram shown in Fig. \ref{fig:sag2} describes the data flow of the SAG system during the observations related to a sub-array with one or more telescopes involved. The \textit{Array Data Handler} acquires the data streams from the telescopes, and the sag-reco receives this data stream. This software component performs the reconstructions of raw data in several steps. The first reconstruction step (DL0 to DL1) is executed for each telescope data stream, while the following steps (DL1 to DL3) merge the data acquired by all telescopes in the sub-array and require the software array stereo trigger (SWAT). The results are input for the next sag-reco steps (from DL1 to  DL3) and the SAG Data Quality pipelines (sag-dq). The sag-dq performs several data quality checks on the different data levels reconstructed by the sag-reco. Some of the data quality checks are performed on the single telescope data, while others are performed after the stereo trigger analysis. Finally, the data quality results of all sag-dq pipelines are aggregated and then stored in a database. These results are also used to check if the data quality level required to generate a candidate science alert are satisfied. The DL3 resulting from the sag-reco pipeline is the input for the High-Level Reconstruction pipeline (sag-sci) that executes scientific analyses on the DL3 with several science tools. First, the DL3 data are collected in a database. Then, the Pipeline Manager, a component of sag-sci, executes the configured analyses in parallel to obtain the DL4 results (counts maps and analysis) stored in a database. During these analyses, the sag-sci can detect candidate science alerts and send them to the \textit{Transients Handler} system to perform further investigations. The operator and the support astronomer can visualise the results of both the sag-dq and sag-sci pipelines using the \textit{operator HMI}, an ACADA system  that shows the results stored in the databases.

Slurm\footnote{https://slurm.schedmd.com/} manages the workload of different pipelines, also \abr{allows} the scaling on a cluster of machines.

Prototyping versions of the SAG, called Real-Time Analysis (RTA), were developed in the past \cite{bulgarelli1, bulgarelli2, vuillaume, bulgarelli}. Lesson learned and experience has been included in the current design.

\subsection{Online Low-Level Reconstruction - sag-reco}

The online Low-Level Reconstruction pipeline (sag-reco)  executes reconstruction in parallels for different sub-array observing simultaneously with a latency of $15 \: s$. The sag-reco pipeline is composed of the following main components:
\begin{enumerate}
    \item Hillas parameters extraction (DL0 to DL1);
    \item Spatial reconstruction, energy estimation and background tagging (DL1 to DL2, \abr{the event list});
    \item Background rejection selection, quality selection and celestial coordinate reconstruction (DL2 to DL3).
\end{enumerate}
	
 DL0 data is received from the \textit{Array Data Handling} and information about telescopes triggered from the sub-array triggering system. The image integration is performed to obtain a camera map of integrated charges and a camera map of mean arrival time (DL1a). A tail-cut cleaning is performed from these maps, and image parameters (first moments of the signal, global intensity, time gradient) are extracted by the Image Parameter Extractor (DL1b). These parameters are fed into a machine learning model trained on Monte-Carlo simulations to reconstruct the physical event parameters (direction, energy and particle type - DL2). The last stage of sag-reco applies selection cuts to produce DL3 data that is then provided to the High-Level Analysis Pipeline (sag-sci). This component is also able, depending on the configuration, to inform ADH if an event should be stored, providing means for further data volume reduction.
 
 Sag-reco is based on hipeCTA software library \cite{vuillaume} for the DL1 production, and on lstchain \cite{ruben} for the DL1->DL3 steps.

\subsection{Online data quality Analysis - sag-dq}

The online Data Quality Software (sag-dq) is a pipeline that performs a real-time quality check of the data acquired with a latency of less than $5 \: s$ with respect to the data provided as input, performing the data quality monitoring of data levels from DL0 to DL3. 
%The online data quality analysis is necessary to assess the instruments’ health during the data acquisition:  this analysis is crucial to confirm good gamma-ray detection before the generation of a candidate science alert from sag-sci.  

For the development of sag-dq, a Python software library, \abr{ called rta-dq-lib \cite{baroncelli}}, to perform the online data quality analysis of CTA data has been developed. This library is dedicated to the rapid prototyping of data quality use cases. The library allows the user to define, through some XML configuration files, the format of the input data and, for each data field, which quality checks must be performed and which types of aggregations and transformations must be applied.  Internally, it translates the XML configuration in a computational direct acyclic graph (DAG) composed of several types of nodes.  This computational structure and the configuration with  XML  files allow the rapid development of data quality pipelines. 

The computational direct acyclic graph is composed of several types of nodes: the DataSource nodes acquire the data from, e.g. hdf5 files, Executor nodes aggregate or transform a data field, QualityCheck nodes monitor a data quality indicator, and output nodes write the results on disk. The Executor node can be stateless (without memory) or stateful. The DAG model encodes the dependencies of the computational tasks to be performed, and it allows the library to take advantage of parallelization at the thread level easily. More details and an example of DAG is reported in \cite{baroncelli}.

\subsection{Online High-Level analysis: science monitoring and candidate science alert generation - sag-sci}

The High-Level online analysis (sag-sci) is designed to perform scientific analysis in real-time during the observations and detect candidate science alerts. This pipeline shall execute analysis in parallels for different sub-array observing at the same time and is developed using the RTApipe framework described in \cite{parmiggiani1}. It shall detect candidate science alerts with a latency of $5 \: s$ and an integral sensitivity within a factor two of that required for the off-line analysis. 

The input data for this component is the DL3 produced by sag-reco, saved into a database and automatically triggers the sag-sci pipeline analysis. Candidate science alert is cross-checked with the result of the sag-dq before issuing them to the \textit{Transients Handler} system. The results of the scientific analyses are stored in a database, and the \textit{HMI} can show this information to the support astronomer.

\subsection{Development status}

The first release of ACADA has been \abr{delivered} in June 2021 and includes the Supervisor, sag-reco and sag-dq.
The overall ACADA system that includes the current version of the SAG sub-system has been tested and integrated into the computing cluster composed of 15 machines for supporting the Assembly, Integration, and
Verification (AIV) of ACADA \cite{antolini} in the
DESY data centre in Zeuthen \cite{murach}. This environment contains a continuous integration setup based on Jenkins\footnote{https://www.jenkins.io/}
and a software quality assurance environment based on SonarQube \footnote{https://www.sonarqube.org/}.

The released version of the sag-reco and sag-dq are also installed at the CTA site in La Palma and are being integrated with the acquisition system
of the first Large Size Telescope (LST) \cite{cortina}. 

An advanced prototype of rta-sci \cite{parmiggiani1}, partially developed for the AGILE Observatory \cite{bulgarelli} will be adapted and integrated into the next ACADA release.

\section{Conclusions}
The \textit{Science Alert Generation} system will perform the online CTA scientific analysis of data acquired from the array of telescopes in both northern and southern sites; \abr{the latency of the system, the required sensitivity, and the complexity of the data analysis  are challenging because of the large data rate, a few Gbit/s to be reduced and analysed in real-time.} 

Thanks to the unprecedented sensitivity that the SAG will achieve, to the technological studies to manage the data rate, and to the optimisation of data analysis algorithms, CTA will be capable of following  \abr{transient} phenomena in real-time. This will provide real-time feedback to external received alerts and issue candidate science alerts, enabling CTA to maximise the science return on \abr{time-variable} phenomena in a multi-wavelength and multi-messenger context.

\acknowledgments
We gratefully acknowledge financial support from the
agencies and organizations listed here: http://www.cta-observatory.org/consortium acknowledgments

\clearpage
%\subfile{authors}

%% Full authors list (ONLY FOR COLLABORATIONS)
%\clearpage
%\section*{Full Authors List: \Coll\ Collaboration}
%
%\noindent \textbf{Note comment afterwards:} Collaborations have the possibility to provide an authors list in xml format which will be used while generating the DOI entries making the full authors list searchable in databases like Inspire HEP. For instructions please go to icrc2021.desy.de/proceedings or contact us under icrc2021proc@desy.de.\\
%
%\scriptsize
%\noindent
%first.author$^1$, 
%second.author$^2$, 
%third.author$^3$ % .... more names
%and 
%last.author$^{n}$ \\
%
%\noindent
%$^1$first.affiliation.
%$^2$second.affiliation. % .... more affiliation
%$^{m}$last.affiliation.


\begin{thebibliography}{99}

\bibitem{patricelli}
B. Patricelli et al. these proceedings

\bibitem{alicia} A. López-Oramas et al., these proceedings

\bibitem{hess} F. Aharonian et al. Astron. Astrophys. 457, 899 (2006), astro-ph/0607333.
\bibitem{magic}  J. Aleksić et al., Astroparticle Physics 72, 76 (2016), 1409.5594.
\bibitem{veritas} J. Holder, in Proceedings SciNeGHE 2006 (2007), pp. 69–76, astro-ph/0611598.

\bibitem{cta} B. S. Acharya et al. (CTA Consortium) (WSP, 2018), ISBN 9789813270084, astro-ph/1709.07997.

\bibitem{fioretti} V. Fioretti, et al., Proceedings of the 36th International Cosmic Ray Conference - Madison, WI, USA (PoS(ICRC2019)673)

\bibitem{bosnjak} Ž. Bošnjak, A. M. Brown, A. Carosi, et al., arXiv e-prints (2021), 2106.03621.

\bibitem{Igor}
I. Oya et al., 17th Int. Conf. on Acc. and Large Exp. Physics Control Systems, 1046-1051

\bibitem{chiozzi} G. Chiozzi, et al., “CORBA-based Common Software for the ALMA project”, in Proc SPIE 4848, 43, 2002.
doi: 10.1117/12.461036

\bibitem{baroncelli}
L. Baroncelli, et al., in Astronomical Data Analysis Software and Systems XXX, ser. Astronomical Society of the Pacific Conference Series, 2021.

\bibitem{parmiggiani1}
N. Parmiggiani, et al., in Astronomical Data Analysis Software and Systems XXX, ser. Astronomical Society of the Pacific Conference Series, 2021.

\bibitem{dipiano}
A. Di Piano et al, these proceedings

\bibitem{bulgarelli1} A. Bulgarelli, et al., in Proc. SPIE 9145,
Ground-based and Airborne Telescopes V, paper 91452X, 2014. doi: 10.1117/12.2054744

\bibitem{bulgarelli2} A. Bulgarelli, et al., in Proceedings of the 34th International Cosmic Ray Conference (ICRC2015), The Hague, The Netherlands. 

\bibitem{vuillaume} T. Vuillaume, et al., in Proc. ICRC’19,
Madison, Wisconsin, USA, 2019.

%\bibitem{ruben} R. Lopez-Coto et al. for CTA LST project, “lstchain: An analysispipeline for LST-1, the first prototype Large-Sized Telescope of CTA,”inAstronomical Data Analysis Software and Systems XXX, ser. Astro-nomical Society of the Pacific Conference Series, 2021.

\bibitem{ruben} R. Lopez-Coto et al. for CTA LST project, in Astronomical Data Analysis Software and Systems XXX, ser. Astronomical Society of the Pacific Conference Series, 2021.

\bibitem{antolini} E. Antolini, D. Melkumyan, K. Mosshammer, and I. Oya, presented at the ICALEPCS'19, New York, NY, U

\bibitem{murach}  T. Murach, et al., in Proc. SPIE 10707, Software and Cyberinfrastructure for Astronomy V, paper 107070D, 2018. 

\bibitem{cortina} J. Cortina, et al., in Proc. ICRC’19, Madison, Wisconsin, USA, 2019.

\bibitem{bulgarelli} A. Bulgarelli et al., Experimental Astronomy 48 (2019) 199.




\end{thebibliography}
\end{document}